# Artificial intelligence in peer review: How can evolutionary computation support journal editors?

Maciej J. Mrowinski[1,*], Piotr Fronczak[1], Agata Fronczak[1], Marcel Ausloos[2,3] Olgica Nedic[4],

[1]Faculty of Physics, Warsaw University of Technology, Koszykowa 75, PL-00-662, Warsaw, Poland
[*]corresponding author

[2]School of Management, University of Leicester, University Road, Leicester LE1 7RH, United Kingdom

[3]GRAPES - Group of Researchers for Applications of Physics in Economy and Sociology, rue de la Belle Jardinière 483, B-4031, Angleur, Belgium

[4]Institute for the Application of Nuclear Energy (INEP), University of Belgrade, Banatska 31b, Belgrade-Zemun, Serbia
Sub-editor in the Journal of the Serbian Chemical Society

**Abstract**

With the volume of manuscripts submitted for publication growing every year, the deficiencies of peer review (e.g. long review times) are becoming more apparent. Editorial strategies, sets of guidelines designed to speed up the process and reduce editors' workloads, are treated as trade secrets by publishing houses and are not shared publicly. To improve the effectiveness of their strategies, editors in small publishing groups are faced with undertaking an iterative trial-and-error approach. We show that Cartesian Genetic Programming, a nature-inspired evolutionary algorithm, can dramatically improve editorial strategies. The artificially evolved strategy reduced the duration of the peer review process by 30%, without increasing the pool of reviewers (in comparison to a typical human-developed strategy). Evolutionary computation has typically been used in technological processes or biological ecosystems. Our results demonstrate that genetic programs can improve real-world social systems that are usually much harder to understand and control than physical systems.

## Introduction

Peer reviewed publications remain the cornerstone of the scientific world [1–5], despite the fact that the review process itself is far from perfect: The authors are frustrated by having to wait for the delaying reports [6, 7] and the editors are irritated by the time-consuming tasks of searching for and rounding up an excessive number of reviewers in the hope of receiving one or two reports. Surprisingly, despite the importance of peer review, it remains an understudied subject [8]. Existing papers concentrate on statistical or ethical issues relating to the process [9–11], with peer review only recently capturing the attention of complex systems scientists [12, 13]. The goal of our work, which is a result of a collaboration within COST action PEERE (New Frontiers of Peer Review; www.peere.org), is to provide quantitative improvements to the efficiency of the review process.

One of the biggest obstacles to research on peer review is the scarce availability of data. Although it is relatively simple to obtain information about citations or publication dates, the inner workings of the editorial process are usually kept secret. However, thanks to the cooperation with our partners at PEERE, we were able to acquire a very detailed dataset from one of the sub-editors of the Journal of the Serbian Chemical Society (JSCS). The dataset contained information about 58 articles submitted to the journal for publication and allowed us to study the review process from the viewpoint of a journal editor. We were able to separate the process into distinct phases (Fig 1a), determine the duration of each phase and the probability of transitions between phases [14]. Most importantly, we were able to create a simulation platform capable of generating artificial review threads (i.e. realizations of the review process). Formally, each review thread represents a path through the review process decision diagram (Fig 1b), with durations assigned to each transition. To our knowledge, most existing peer review simulations are based on abstract models. In contrast, our approach allows for data-driven simulations, which may lead to a quantifiable improvement in the efficiency of the peer review process.

## Materials and methods

Editorial workflows usually comprise many time-consuming tasks. First, editors need to find appropriate reviewers for each manuscript. Although this can be done with the help of various scientific databases, it is still a very laborious process. Second, they send invitations, handle the communication with reviewers and evaluate the received reports. Finally, they have to deal with situations where they do not receive enough reviews from the invited reviewers or the reviewers give conflicting recommendations.

Before we delve into a more in-depth explanation of how editorial workflows can be optimized, we would like to take a closer look at the artificial review threads. A realization of the review process is initiated when the editor issues an invitation to a potential reviewer. Thus, to mirror the behavior of real-life realizations, artificial review threads begin their life in the INVITATION phase (Fig 1). The next phase is determined by transition probabilities calculated from data. Unfortunately, the most probable route (73%) leads to the INQUIRY phase, which corresponds to the situation in which the invited reviewer does not answer. This prompts the editor, who waits for seven days after sending the initial invitation, to send another e-mail to the reviewer (the inquiry). When this second e-mail remains unanswered for ten days, then the editor assumes that the process is over with a negative outcome (without a review), which corresponds to the artificial review thread ending in the NO RESPONSE phase. However, if the reviewer does actually answer the inquiry (or the initial invitation) and agree to write the review, then the review thread enters the CONFIRMATION phase. In such a case, the editor waits for twenty-five days for the review and sends an inquiry if it is not received. To sum it up, each artificial review thread can end in one of three states – REPORT (which is the desired outcome), NO RESPONSE, and REJECT (when the invited reviewer answers but refuses to write the review). The probabilities of all outcomes, as well as probabilities of transitions between intermediate phases and durations of these transitions, were calculated using JSCS data.

To make the simulations as realistic as possible, all the review threads strictly followed the decision diagram presented in Fig 1a. After the initial invitations, the system handled all the interactions with the reviewers using real-world data. The remaining part of the review process was labelled 'editorial strategies'. The editorial strategies are sets of rules that tell the editors how many new invitations they should issue at the beginning of the process and how many they should issue each time one review threads ends.

Informal discussions with editors revealed two editorial strategies, both of which are visualised in Fig 2. According to the first strategy (for simplicity called the strategy of Editor A), the editor should start a set number of review threads (called the batch size) at the beginning of the process and wait for all the threads to end before repeating this procedure if the required number of reviews is not received. In all of our simulations, the batch size is a parameter which, as in the strategy of Editor A, determines the upper limit of the number of simultaneously running review threads. The second strategy (for simplicity called the strategy of Editor B) begins in a similar way, with the editor sending invitations to a batch of reviewers. The difference between Editor A and Editor B is that the latter is always keeping the entire batch running. A finished review thread is immediately replaced with a new one, which ensures that the total number of running threads is equal to the batch size at all times.

In determining which strategy is better, one should consider that two opposite forces affect the efficiency of editorial strategies. On the one hand, both authors of articles and editors would like to receive reviews as soon as possible. On the other hand, the editors prefer to minimize the number of invitations sent to reviewers. Studies show that authors become impatient if the review process takes too long and may withdraw the manuscript [15]. In theory, simply inviting many reviewers to assess the manuscript could shorten the review time. However, this is not a feasible solution for editors, as there are costs (e.g. searching time) associated with each reviewer. It follows that the best editorial strategy is the one that achieves the shortest review time using the smallest number of reviewers.

Taking the aforementioned factors into account, the strategies can be compared using the simple diagram presented in Fig 3a, which shows the review time as a function of the effective number of reviewers for both strategies. The effective number of reviewers is simply the average number of reviewers required by a particular strategy in order to acquire two reviews per article. In general, the effective number of reviewers will be higher than the batch size needed to achieve a given review time. As can be clearly seen, the strategy of Editor A is better than the strategy of Editor B. This may seem counter-intuitive at first, as the review time of Editor B for the same batch size is shorter than that of Editor A (Fig 3c). However, the effective number of reviewers that Editor B employs is larger (Fig 3d), which ultimately makes the strategy of Editor B less efficient.

The strategies of Editor A and Editor B are simple and intuitive. Many new editors would likely adopt one of these strategies as the first go-to solution in the peer review process. However, editors, especially in smaller journals, usually organize their workflows on their own without any point of reference. The development of an automated tool could provide more efficient strategies, thereby making the editor's job much easier, shortening review times and increasing the satisfaction of authors of manuscripts with the review process.

Cartesian Genetic Programming (CGP; implementations in various programming languages are readily available at www.cartesiangp.co.uk) [16–18] was originally conceived as a way of designing digital circuits. However, the actual scope of its applications is much larger, and it may be ideal for use in the optimization peer review process. At the core of CGP is an evolutionary algorithm, which creates computer programs (functions) using concepts from the theory of evolution (i.e. natural selection and mutation) by optimizing some fitness function. In the present case, these programs corresponded to editorial strategies, which told the editor how many new review threads should be started at any given time.

The editorial strategies in CGP can be encoded as a grid of nodes (Fig 4). Leftmost nodes represent inputs to the strategy – this is where the editor provides the integer numbers that describe the system (the required number of reviews, the number of received reviews, the number of running review threads, and the size of the batch) at a particular time. Input nodes are followed by intermediate nodes —a grid of $n$ rows by $m$ columns filled with nodes that correspond to basic mathematical functions. Each node has two inputs, which are used to calculate the value of the function realized by the node, and one output. It should be emphasized that inputs of nodes in a given column can be connected only to outputs of nodes in preceding columns. Forward (recurrent) connections are not allowed in order to avoid loops. In our simulations, we used only a single row and two thousand columns—that is, there were two thousand intermediate nodes between input and output arranged on a single, one-dimensional line. From our experience, this number of nodes is a good starting point as having too few nodes effectively prevents optimization (CGP is not able to find a working solution to the problem in a reasonable time). All functions operated only on integers and were protected, which means that for inputs that are not allowed (e.g. division by zero) they would always return the number one. The rightmost node in the strategy is the output node, which by connecting to one of the preceding nodes tells the user of the strategy where to look for the final value.

Editorial strategies were evolved using a very simple algorithm, called the '4 + 1 evolutionary algorithm:' [18]

1. At the beginning of the simulation five strategies are created by randomly assigning functions and links (connections) to nodes. Each of these random strategies is evaluated using a fitness function and the best strategy is chosen as the parent – the seed of the simulation.
2. At each step of the simulation the current parent is copied four times and the mutation operator is applied to these copies. The mutation operator traverses all inputs of all nodes in a strategy and changes them randomly with a given probability (called the mutation probability). It also traverses all function realized by nodes and changes them randomly with the same probability. It means that both the function and inputs can be changed by the mutation operator at the same time. Then, each of the mutated copies is evaluated using the fitness function. If one of the copies is at least as good (fit) as the parent, then it becomes the new parent. Otherwise, if all copies are worse strategies than the parent, then they are discarded and the parent carries over into the next iteration. We tested mutation probability values in the range $p_{mut} \in [0.01, 0.10]$. Since we were able to reach optimized solutions for all probabilities in this range, we decided to use $p_{mut} = 0.03$ in all simulations. It is worth mentioning at this point that while a graph representing an editorial strategy may be comprised of a large number of nodes, usually only a small fraction of these nodes (<10%) are actually used to calculate the output value. That is, most of the nodes in the graph are not parts of paths that connect inputs with the output. We call such nodes *inactive*. When the mutation operator is applied to an inactive node, the genotype may change but the phenotype remains the same. On the other hand, applying the mutation operator to an active node may change some of the previously inactive nodes into active as a result of changing the paths that lead from inputs to the output. This gradual build-up of changes in the inactive parts of graphs is one of the most important features of CGP.
3. Step 2 is repeated until a strategy characterized by a sufficiently low fitness function value is found (when compared to real strategies developed by human editors). There is no natural stopping condition for this algorithm – in theory it may try to optimize the strategy indefinitely.

The fitness function used to evaluate strategies is one of the most important components of the optimization process. The numerical value it assigns to each program is a measure of effectiveness and can be used to compare two strategies. Our approach is based entirely on real-world data. With the help of artificial review threads, the fitness function we designed performs numerical simulations of the review process for batch sizes between two and twenty. The effective number of reviewers and the review time are calculated by averaging the results of simulations of the review process of ten thousand articles per one batch size. The pseudo-code for the fitness function and the function responsible for the simulation of the review process can be found in the Supporting Information (S1 Algorithm).

During simulations we assume that time is discrete and that one time step corresponds to one day. However, we also assume editors send new invitations (and consult editorial strategies) only when one of the running review threads finishes its execution (that is, when it enters the NO RESPONSE, REPORT or REJECTION phase). A valid strategy must obey certain constraint. First, it is assumed that at the beginning of the review process (i.e. the zero-th day) invitations are sent to the entire batch of reviewers. Second, the batch size must never be exceeded. Third, at least one invitation must be sent when there are no active (running) review threads. If at least one of these constraints is not met by the strategy being evaluated (the first condition is imposed automatically during simulations), then the fitness function returns a very large value (proportional to the number of batch sizes for which the strategy is not valid), unobtainable by a proper strategy. Otherwise, when the aforementioned conditions are fulfilled, the fitness function calculates the area under the strategy efficiency plot (Fig 3; the integral is calculated over the interval [7, 15]) and returns that value to the CGP algorithm. Despite the apparent simplicity of this definition of the fitness function, it is entirely data driven and leads to improvements in real-world strategies. Lastly, it should be noted that we did not use any well defined stop condition. We would wait until the value of the fitness function indicated that the evolved strategy was better than the strategy of Editor A and we would terminate the optimization process manually once it reached a stable solution (the value of the fitness function stopped changing).

Although the simulations performed to evaluate the values of the fitness function were based on 'full' review threads (i.e. with full information about phases of the process that constitute each thread), the results of these simulations can be recreated using data presented in the Supporting Information – S1 and S2 Tables. Both tables contain empirical distributions of the durations of the review process in JSCS. The duration, expressed in days, can be found in the 'Days' column. The 'Freq.' column contains the number of review threads in the JSCS sample that correspond to each duration. We should add at this point that the journal agreed to the publication of the data in this form. Also, the data we received was completely anonymized; we did not have access to any personal information.

## Results and discussion

With the help of CGP equipped with the aforementioned fitness function, we searched for evolved strategies that would be better than strategies conceived by humans. The evolved strategy can be visualized in a couple of ways. The most natural representation is presented in Fig 4, which depicts the optimal strategy we found as a graph. It is worth mentioning that this is not the full program that was initially evolved by CGP. In order to simplify the graph, we employed CGP once again—this time to look for a new strategy which would be functionally equivalent to the evolved strategy but implemented using a smaller number of active nodes (that is, a strategy with the same phenotype but simpler genotype). Also, to make it more readable, we removed all inactive nodes from Fig 4 – that is nodes that are not directly or indirectly connected to the output and thus do not take part in the calculation of the value returned by the strategy. For the sake of comparison, S1 Fig from Supporting Information shows an extract of the full strategy graph with inactive nodes intact.

Although the program in Fig 4 looks complicated, the strategy it represents is not. For example, if the batch size is four, then the editor should initially invite two reviewers. The editor should then wait until only one active thread remains. If only one review has been received up to that point, the editor should continue to wait and begin two new review threads only if the last remaining thread does not yield a review. Otherwise, the editor should send invitations to two new reviewers immediately, without waiting for that last thread.

Since editorial strategies are essentially deterministic functions that assign integer numbers to states of the system (to each possible input), one can recreate the mathematical formula that corresponds to a strategy by following nodes from output to inputs and writing down the nested functions encountered in nodes. It is important to keep in mind that all functions realized by nodes are protected and normal rules of simplifying mathematical expressions often do not apply. The function equivalent to the evolved strategy can be expressed in the following way:

$$N_{NRT} = [N_R \pmod{N_{NR}} + N_{EQ} \pmod{N_{RR}} / N_{ART}] [[N_B - N_{ART}] / N_{RQ}] [N_{RQ} \pmod{N_{RR}} / N_{ART}]$$

where $N_{NRT}$ is the number of new review threads that should be started, $N_{RQ}$ is the required number of reviews, $N_{RR}$ is the number of received reviews, $N_{ART}$ is the number of active review threads, and $N_B$ is the batch size.

The evolved strategy performed well when compared to that of Editor B and Editor A (Fig 3). For eight effective reviewers, the review time was 17 days shorter using the evolved strategy as compared to the strategy of Editor B and six days shorter as compared to the strategy of Editor A. Considering that the review time of the JSCS is relatively short, editors of other journals can potentially shorten their peer review times even more. We do not claim that the strategy presented in Fig 4 is universal. It works in the ecosystem of the JSCS but may be less efficient in a different context. However, the method used is universal. As long as editors are able to perform a statistical analysis of the peer review process of their journal and use the results to create simulations of the process with artificial review threads, they can employ CGP to improve their workflow. We believe the work presented herein is an interesting application of an evolutionary technique to what, in part, is a sociological problem and that such a data-driven approach is the key to realistic models of peer review that can provide more insight into the process and make it more efficient.

It should be noted that our method of optimization has no bearing on the quality of peer review. Editorial strategies determine when and how many invitations must be sent to reviewers but they do not interfere with the actual process of reviewing the manuscript by said reviewers. It means that it is still up to the editor to uphold the standard of the journal and the quality of the process.

We would like to end this article by mentioning one more discovery. Editorial workflows are incredibly important, but what can scientists do to speed up the review process? Surely, not everything is in the hands of editors. As it turns out, scientists can do quite a lot. Fig 3 shows what we call 'the perfect world strategy.' In order to create the efficiency curve for this strategy, the evolved strategy was evaluated under a very simple, yet unrealistic assumption, as shown by data: All potential reviewers answer their mails and always immediately respond to an initial invitation by either accepting or refusing to write the review. That is, the NO RESPONSE and REJECTION outcomes of the review process are merged into one without changing the probability of accepting to write a review. As can be clearly seen, this one assumption makes a striking difference. In the end, it seems that a shorter review time is easily obtainable. It is, quite literally, one mouse click away.

## Acknowledgments

This publication is supported by the COST Action TD1306 "New frontiers of peer review".

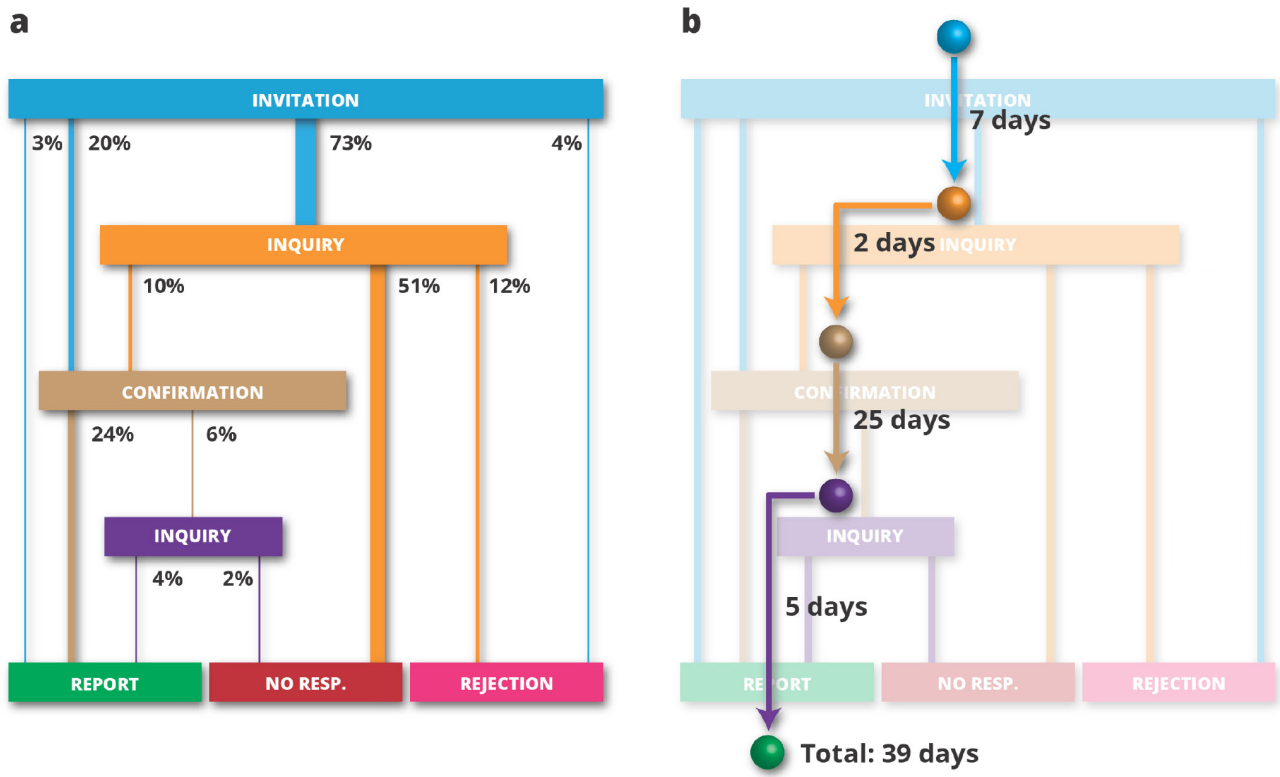

Figure 1: Decision diagram of the review process. **a**, Transition probabilities - the percent of review threads that pass through the edge. **b**, An example of a review thread - a path through the decision diagram coupled with the duration of each phase.

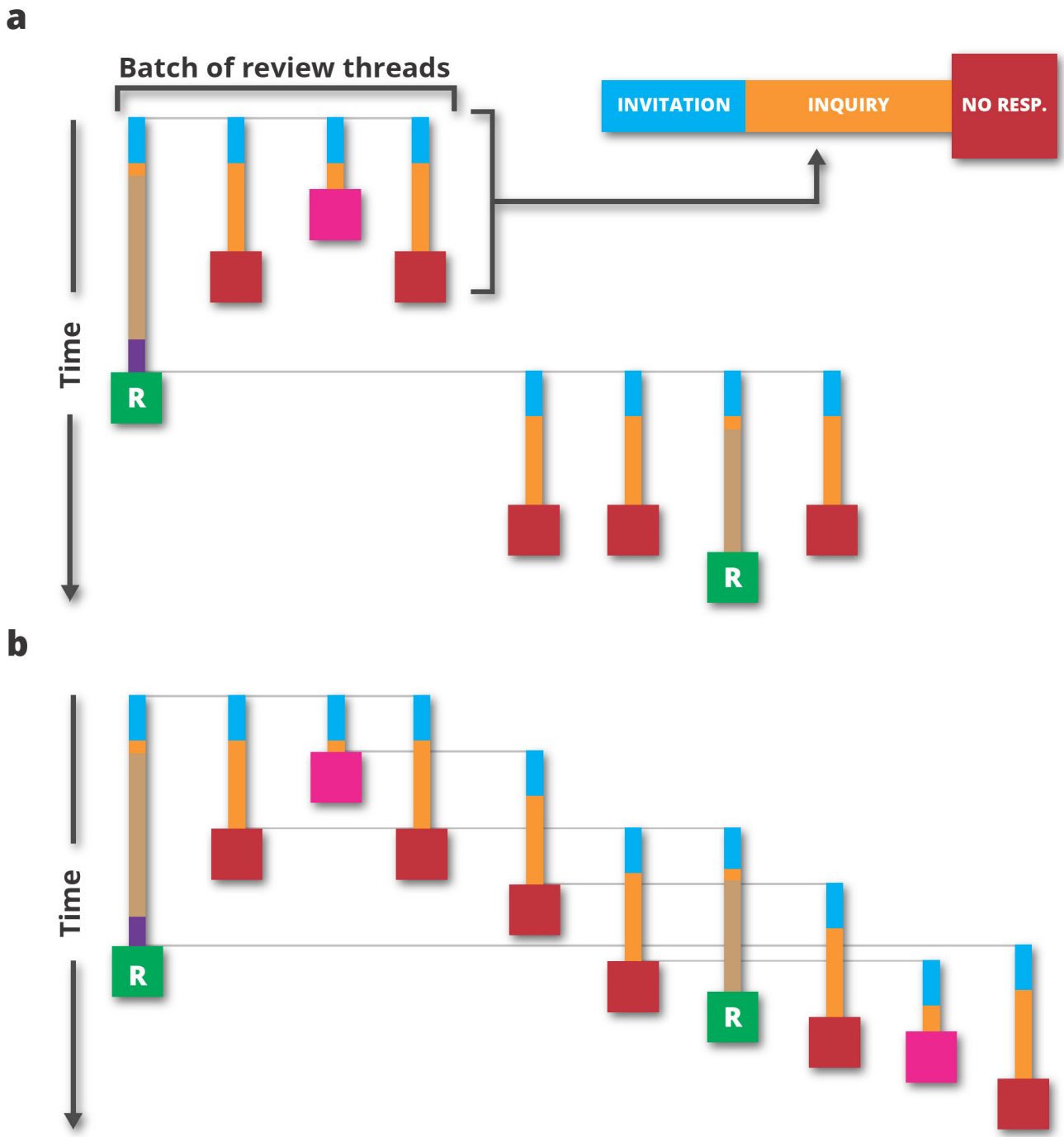

Figure 2: Two strategies popular among editors. **a**, Strategy of Editor A, according to which one should wait for all review threads to end before starting new ones. **b**, Strategy of Editor B, where the number of active review threads is always constant - a new invitation is sent immediately after one of the active threads ends.

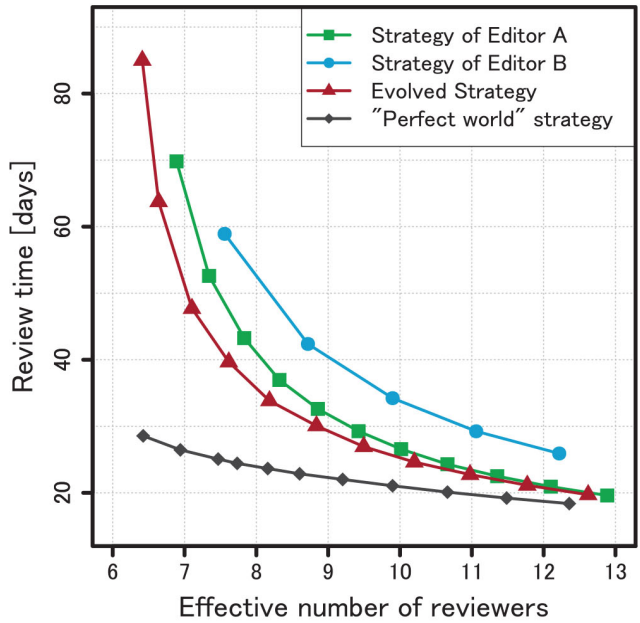 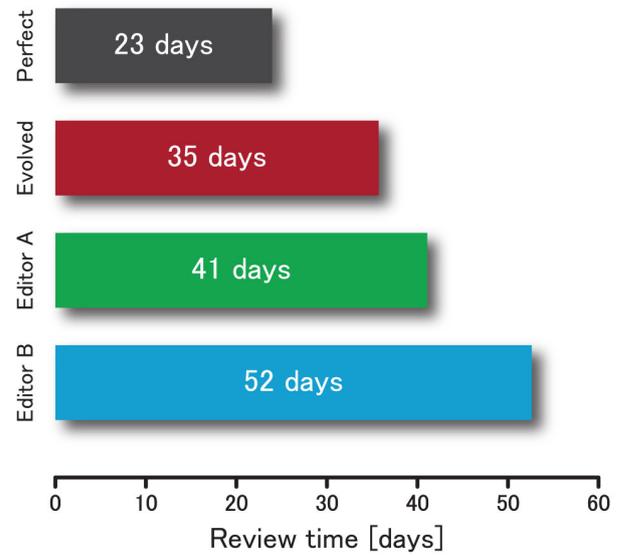

Figure 3: Efficiency of strategies. Strategies were evaluated by simulating the review process of a large number of articles (using the same procedure that was employed in the fitness function) and while the process itself is stochastic, deviations from these efficiency curves are negligible. (a) Review time as a function of the effective number of reviewers (the average number of reviewers needed to achieve a given review time). Each point on the plot corresponds to a single batch size—the first point on each curve represents the batch of two reviewers and for subsequent points the batch size increases by 1. (b) Comparison of the review time of various strategies for eight effective reviewers.

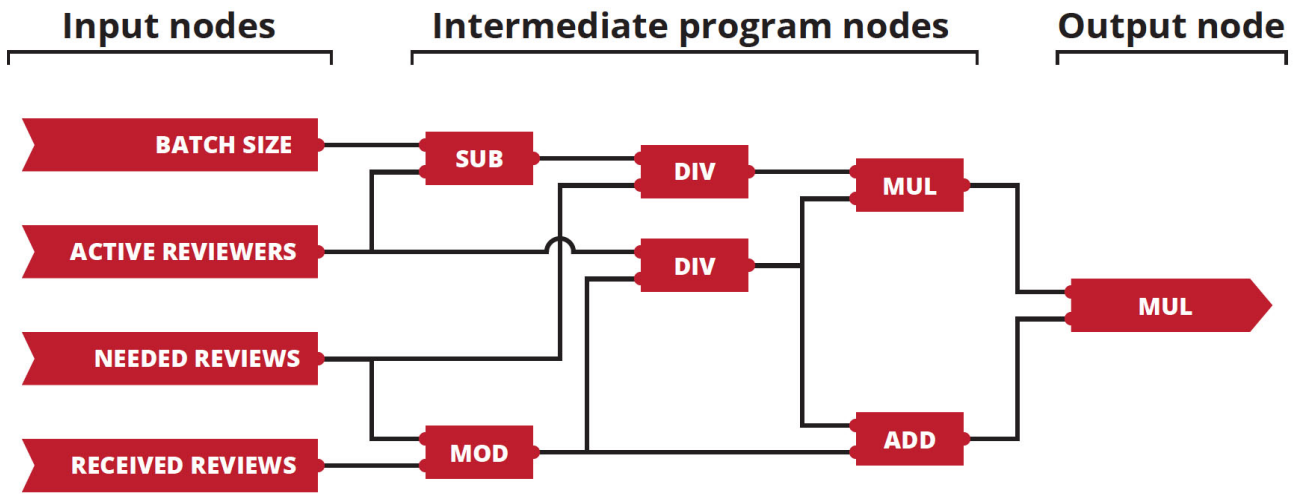

Figure 4: Evolved editorial strategy. BATCH SIZE, the size of the batch; ACTIVE REVIEWERS, the number of active review threads (reviewers for whom the review process has not ended yet); NEEDED REVIEWS, the number of reviews required per article; RECEIVED REVIEWS, the number of reviews received thus far; SUB, subtraction; DIV, division; MUL, multiplication; ADD, addition; MOD, modulo.

**S1 Table.** Distribution of the length of the review process (single review thread; process ended with a review) in JSCS.

| Days | Freq. | Days | Freq. | Days | Freq. | Days | Freq. |
|---|---|---|---|---|---|---|---|
| 0 | 1 | 11 | 1 | 22 | 4 | 33 | 2 |
| 1 | 1 | 12 | 4 | 23 | 6 | 34 | 3 |
| 2 | 5 | 13 | 3 | 24 | 6 | 35 | 3 |
| 3 | 3 | 14 | 4 | 25 | 3 | 36 | 3 |
| 4 | 2 | 15 | 4 | 26 | 4 | 38 | 1 |
| 5 | 2 | 16 | 3 | 27 | 3 | 39 | 3 |
| 6 | 3 | 17 | 6 | 28 | 6 | 40 | 2 |
| 7 | 4 | 18 | 3 | 29 | 4 | 41 | 3 |
| 8 | 4 | 19 | 3 | 30 | 2 | 53 | 1 |
| 9 | 1 | 20 | 6 | 31 | 2 | 55 | 1 |
| 10 | 3 | 21 | 4 | 32 | 3 | 68 | 1 |

**S2 Table.** Distribution of the length of the review process (single review thread; process ended without a review) in JSCS.

| Days | Freq. | Days | Freq. |
|---|---|---|---|
| 1 | 5 | 11 | 1 |
| 2 | 1 | 16 | 4 |
| 3 | 2 | 17 | 89 |
| 4 | 1 | 18 | 20 |
| 8 | 14 | 37 | 1 |
| 9 | 14 | 40 | 7 |
| 10 | 15 | 41 | 1 |

[S1 Algorithm.](#) Algorithms for the fitness function and the simulation of the review process.

**Algorithm S1a:** The algorithm which evaluates the fitness of editorial strategies.

```
Function fitness(editorialStrategy: editorial strategy, minBatch: integer, maxBatch: integer,
minEffective: integer, maxEffective: integer, plannedSimulationRuns: integer, criticalFitness: integer):
    points ← {};                                              // initialise empty list of Point objects
    criticalErrors ← 0;
    foreach batchSize between minBatch and maxBatch do
        point ← (0, 0);                                       // initialise Point point to (x = 0, y = 0)
        simulationRuns ← 0;
        repeat
            simulationRuns ← simulationRuns + 1;
            elapsedDays, effectiveReviewers, criticalError ← simulation(batchSize, editorialStrategy);
            if criticalError = true then
                criticalErrors ← criticalErrors + 1;
                break
            else
                point.x ← point.x + effectiveReviewers;
                point.y ← point.y + elapsedDays;
            end
        until simulationRuns < plannedSimulationRuns;
        point.x ← point.x/simulationRuns;                     // averaging
        point.y ← point.y/simulationRuns;
        points ← points ∪ {point} ;
    end
    if criticalErrors > 0 then
        return criticalErrors * criticalFitness ;             // penalty for errors in strategy
    else
        return the area under the curve defined by points (points are interpolated by lines; area is
        calculated for x ∈ [minEffective, maxEffective]; if the range of points is smaller, it is assumed
        that the y value of the missing points is equal to the y value of the nearest point in points);
    end
end
```

**Algorithm S1b:** The algorithm used to simulate the review process.

```
structure ReviewThread{
    integer duration;                              // duration, in days, of this review thread
    boolean hasReview;    // indicates whether a review was received during the execution of this
                          thread
    integer offset ← elapsedDays;                  // number of days after which the thread was started
};
Function simulation(batchSize: integer, editorialStrategy: editorial strategy):
    T ← { ReviewThread t_i | i = 1, 2, ..., batchSize};        // generate initial review threads
    threadsNumber ← batchSize;                                  // initial number of threads
    receivedReviews ← 0;
    elapsedDays ← 0;
    effectiveReviewers ← batchSize;                             // initial number of reviewers
    while receivedReviews < 2 do
        elapsedDays ← min{t_i.offset + t_i.duration : t_i ∈ T}; // find the smallest number of days after
        which at least one of the review threads ended
        foreach (t_i ∈ T | t_i.offset + t_i.duration = elapsedDays) do
            if t_i.hasReview = true then
                | receivedReviews ← receivedReviews + 1
            end
            T ← T \ t_i;                          // remove the finished thread from the list of threads
            threadsNumber ← threadsNumber − 1;
        end
        if receivedReviews < 2 then
            newThreadsNumber ← editorialStrategy(state parameters);    // the strategy proposes a
            number of new threads that should be started based on available information
            if newThreadsNumber < 0 or
            newThreadsNumber + threadsNumber > batchSize or
            (threadsNumber = 0 and newThreadsNumber = 0) then
                | return (criticalError ← true);
            else
                T ← T ∪ { ReviewThread t_i | i = 1, 2, ..., newThreadsNumber};  // create new review
                threads
                effectiveReviewers ← effectiveReviewers + newThreadsNumber;
                threadsNumber ← threadsNumber + newThreadsNumber;
            end
        end
    end
    return elapsedDays and effectiveReviewers;
end
```

**S1 Fig.** The first 224 nodes of an editorial strategy with both active (black connections) and inactive (grey connections) nodes visible. The color of a node corresponds to the function realized by the node: light blue - input, red - subtraction,

**green - division, blue - multiplication, pink - addition, yellow - modulo.**

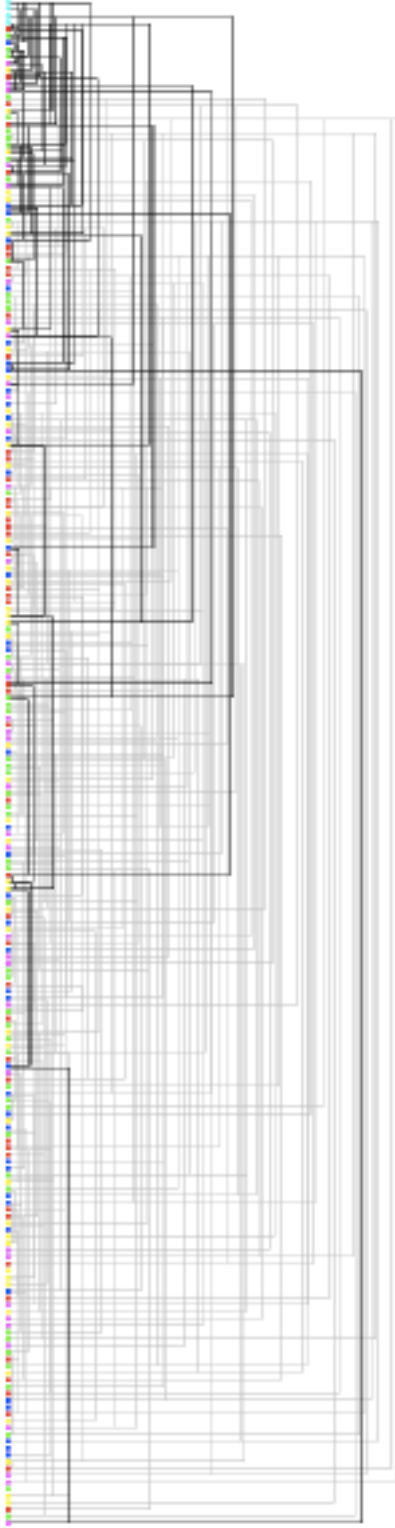